\documentclass[a4paper,12pt]{article}
\pdfoutput=1 
\usepackage{jheppub} 
\usepackage[T1]{fontenc} 
\usepackage{epsfig}
\usepackage{color}
\usepackage{float}
\def\beq{\begin{equation}}
\def\eeq{\end{equation}}
\def\beqn{\begin{eqnarray}}
\def\eeqn{\end{eqnarray}}

\def\met{E_T \hspace*{-1.1em}/\hspace*{0.5em}}
\def\gl{\tilde{g}}
\def\neu{\tilde{\chi}_1^0}

\title{Quark-gluon discrimination in the search for gluino pair production at the LHC}

\author[a]{Biplob Bhattacherjee,}
\author[b]{Satyanarayan Mukhopadhyay,}
\author[c,d]{Mihoko M. Nojiri,}
\author[e]{Yasuhito Sakaki}
\author[f]{and Bryan R. Webber}

\affiliation[a]{Centre for High Energy Physics, Indian Institute of Science, Bangalore, India}
\affiliation[b]{PITT-PACC, Department of Physics and Astronomy, University of Pittsburgh, PA 15260, USA}
\affiliation[c]{Kavli IPMU (WPI), The University of Tokyo, Kashiwa, Chiba 277-8583, Japan}
\affiliation[d]{KEK Theory Center and Sokendai, Tsukuba, Ibaraki 305-0801, Japan}
\affiliation[e]{Department of Physics, Korea Advanced Institute of Science and Technology, 291 Daehak-ro, Yuseong-gu, Daejeon 34141, Republic of Korea}
\affiliation[f]{Cavendish Laboratory, J.J. Thomson Avenue, Cambridge, UK}

\emailAdd{biplob@cts.iisc.ernet.in}
\emailAdd{satya@pitt.edu}
\emailAdd{nojiri@post.kek.jp}
\emailAdd{sakakiy@post.kek.jp}
\emailAdd{webber@hep.phy.cam.ac.uk}

\preprint{\\Cavendish-HEP-16/16,\\IPMU16-0146,\\KEK-TH-1936,\\PITT-PACC 1609}
\abstract{We study the impact of including quark- and gluon-initiated jet discrimination in the search for strongly interacting supersymmetric particles at the LHC. Taking the example of gluino pair production, considerable improvement is observed in the LHC search reach on including the jet substructure observables to the standard kinematic variables within a multivariate analysis. In particular, quark and gluon jet separation has higher impact in the region of intermediate mass-gap between the gluino and the lightest neutralino, as the difference between the signal and the standard model background kinematic distributions is reduced in this region. We also compare the predictions from different Monte Carlo event generators to estimate the uncertainty originating from the modelling of the parton shower and hadronization processes.} 

\begin{document} 
\maketitle
\flushbottom
\section{Introduction}
The current LHC search for strongly interacting supersymmetric particles in a multi-jet final state primarily relies on kinematic discriminants to separate the signal from very large standard model (SM) backgrounds~\cite{ATLAS_8TeV,ATLAS_13TeV_ICHEP,CMS_8TeV,CMS_13TeV}. The signal from heavy squarks or gluinos decaying to a light neutralino lies in the high visible and missing momentum tail. The hadronic jets in the supersymmetry (SUSY) signal come from the decay of a gluino or a squark to one or more quarks and a neutralino. On the other hand, for the dominant SM background of a single weak boson and multiple jets, the jets originate from the initial state radiation of both quarks and gluons. A natural question therefore is whether the difference in the substructure of the decay jets in the signal process and the radiation jets in the SM background can be utilized to further improve the searches. This difference is related to the discrimination of quark- and gluon-initiated jets~\cite{Schwartz1,Schwartz4,Schwartz2,Schwartz3,Larkoski1,ASJ_paper,Larkoski2,Sakaki:2015iya,Badger:2016bpw,FerreiradeLima:2016gcz,Krauss,Moult}, a topic being actively explored by the ATLAS and CMS collaborations~\cite{ATLAS_qg,CMS_qg,ATLAS_qg_8TeV}. 

One goal of the studies in this direction by ATLAS and CMS is to derive template distributions from data for observables that can separate quark- and gluon-initiated jets~\cite{ATLAS_qg,CMS_qg,ATLAS_qg_8TeV}. Such a data-driven approach can avoid uncertainties coming from the Monte Carlo (MC) modelling of the low energy hadronization component, and to a lesser extent of the parton shower implementation. Although the data based templates are still in an early stage of development, especially when employing multiple observables, which requires large statistics not yet reached at the LHC, it is worthwhile to briefly discuss the method and its comparison to MC predictions. The only input from MCs in this approach is the quark and gluon-initiated jet fractions in two different processes~\footnote{If more than two processes can be used, one has a cross-check on the results~\cite{ATLAS_qg_8TeV}.}, for example, in the dijet and $\gamma+$jet or $Z+$jet process, computed at the Born level including parton shower effects. With this definition, depending upon the jet transverse momenta, the dijet event sample consists of $50-60\%$ gluon-initiated jets, while the $\gamma+$jet or $Z(\rightarrow e^+ e^-)+$jet events contain $70-80\%$ quark-initiated jets. The observed normalized distribution in the data for a given observable in the two samples can then be used to derive the normalized distribution for a ``pure'' quark and gluon jet  by solving a pair of linear equations in two variables (for each bin of the normalized distribution, and repeated for different transverse momenta and rapidity intervals). While the uncertainties coming from the parton distribution functions and MC implementation of the Born process and initial and final state radiation are small, the largest systematic uncertainty for such studies arises from the dependence of the templates on the processes being used to derive it~\cite{ATLAS_qg_8TeV}. This key aspect of process dependence requires further studies before data-driven templates can be employed in the long run for physics searches without bringing in additional large systematics. 

Comparison of the templates derived from data and from the MCs following the same procedure described above shows that while the MC predictions for quark-initiated jets agree reasonably well with the distributions extracted from data, the distributions for the gluon-initiated jets differ~\cite{ATLAS_qg,CMS_qg,ATLAS_qg_8TeV}. The data based templates for gluon jets fall in most cases in between the predictions of different MCs ({\tt Pythia}~\cite{Pythia,Pythia8} and {\tt Herwig}~\cite{Herwig} to be specific).  Such difference in MC prediction for the distribution of quark-gluon tagging observables have also been observed in phenomenological studies~\cite{Schwartz3,Larkoski1,ASJ_paper,Larkoski2,Sakaki:2015iya,Badger:2016bpw,FerreiradeLima:2016gcz,Moult}. With this in mind, the usefulness of quark-gluon discrimination in physics searches at the LHC can be studied using existing event generators, and future use of data-driven templates is expected to lead to a performance somewhere in between the {\tt Pythia}- and {\tt Herwig}-based predictions. If promising improvements are found irrespective of the MC used, and after folding in additional systematic uncertainties in the background prediction from substructure variables, the physics case to pursue quark-gluon discrimination as a tool to find new physics at the LHC would be well motivated. 

The goal of this study is to evaluate the expected improvement in the search for gluino pair production at the LHC by including the quark-gluon tagging observables to the standard supersymmetry search strategy in the multijet and missing transverse momentum channel. After including initial and final state parton shower effects to leading order matrix elements, it is estimated that while the third and fourth highest transverse momentum jets in gluino-pair events are expected to be quark-initiated, in the dominant $V+$jets ($V=Z,W$) backgrounds, they are more likely to be gluon-initiated. This leads to a considerable improvement in the signal to background ratio, when jet substructure based observables are utilized. Moreover, including both the kinematic and the jet substructure observables within a multivariate analysis is found to enhance the search prospects further, especially when the mass difference between the gluino and the neutralino lies in an intermediate region. The projected improvement over standard kinematics based searches is observed independent of the MC generator used, though to a different degree. 

In Sec.~\ref{setup}, we describe the quark-gluon separation variables used to define a multivariate discriminant, our Monte Carlo simulation of the signal and background processes as well as the kinematic selection of the signal region. We begin Sec.~\ref{results} by first describing the expected quark-gluon fraction of jets in the signal and background processes based on truth level MC information. This is followed by a discussion on the distribution of relevant kinematic variables. The multivariate analysis procedure is described next, followed by the results on the boosted decision tree based separation of the signal and background jet substructure. Combining the information from both kinematics and jet substructure we obtain the signal and background likelihood distributions, which are then used to estimate the expected LHC search reach using different methods in the gluino-neutralino mass plane. We summarize our findings in Sec.~\ref{summary}.

\section{Analysis Setup}
\label{setup}

\subsection{Overview of quark-gluon tagging variables}
\label{sec:qg_vari}
Based on the difference in splitting probabilities in a parton-shower picture, different possible variables have been proposed for quark-gluon discrimination, which essentially rely on the fact that a gluon produced with similar kinematics leads to a larger multiplicity of soft emissions compared to a quark, and a gluon-initiated jet is wider than a quark-initiated one. These differences follow from the higher colour charge-squared of the gluon, $C_A=3$, versus $C_F=4/3$ for a quark. As demonstrated in previous studies, based on both perturbative methods as well as MC simulations, it is found that the following variables lead to a better quark-gluon separation:
\begin{enumerate}
\item The number of charged tracks inside the jet cone ($n_{\rm ch}$), with each charged track having $p_T>1$ GeV, where $p_T$ denotes its transverse momentum. Even though it is difficult to model this observable accurately by MC generators, the recent ATLAS studies on the charged track multiplicity distribution using $8$ TeV LHC data shows reasonable agreement for a set of MC tunes upto very high jet transverse momenta~\cite{ATLAS_charged}. We shall utilize such tunes in our study for both {\tt Pythia} and {\tt Herwig} MCs, as discussed in Sec.~\ref{MC}. 
\item Energy-energy-correlation (EEC) angularity~\cite{Larkoski1} variables, for example, the observable denoted by $C_1^{(\beta)}$ can be defined in terms of the charged track momenta as
\begin{equation}
C_1^{(\beta)}=\frac{\sum_{i} \sum_{j} p_{T,i} \times p_{T,j} \times (\Delta R(i,j))^\beta}{(\sum_{i} p_{T,i} )^2}.
\end{equation} 
Here, the sums over $i$ and $j$ run over all the tracks associated to the jet with $j>i$, while $\beta$ is a tunable parameter. As determined in previous studies~\cite{Larkoski1}, from perturbative calculations and MC simulations,  $\beta=0.2$ is found to be an optimal choice that maximizes the quark-gluon separation. The distance in the rapidity-azimuthal angle plane between the tracks $i$ and $j$ is denoted by $\Delta R(i,j)$.
\item Jet mass ($m_J$) scaled by its transverse momentum $m_J/p_{T,J}$. 

\item In addition to the above set of variables, as discussed in our previous study~\cite{ASJ_paper}, the input for the number of softer reconstructed jets (associated jets) around a primary hard jet can also improve quark-gluon separation, since it captures additional information from radiation outside the jet radius not included in the above variables. 
\end{enumerate}

In this study we shall use $n_{\rm ch}$, $C_1^{(\beta)}$ and $m_J/p_{T,J}$ as the inputs to a multivariate discriminant for quark- and gluon-like jets. While the inclusion of associated jets can be helpful, it is challenging to do so in a multijet environment, as one needs to remove overlap with ISR jets. We leave the investigation of such overlap removal methods to a future study.

\subsection{Kinematic selection of signal region}
The ATLAS and CMS searches define multiple signal regions determined in terms of kinematic selection criteria that can separate a SUSY squark or gluino production process from the SM backgrounds in the multijets$+\met$ channel. Even though this is a challenging analysis in an hadronic environment, for high squark-gluino masses the hard scale of the signal process is higher than the hard scale of most SM processes. This latter fact is reflected in the high values of sum of jet transverse momenta ($H_T$) or effective mass ($M_{\rm eff}=H_T + \met$) demanded in the signal regions.  Following the ATLAS search strategies for $14$ TeV LHC~\cite{ATLAS14}, we first make a pre-selection of events based on the following cuts:

\fbox{{\bf Cut-1:}}
\begin{enumerate}
\item The number of jets, $n_j \geq 4$, with $p_T^{j_1}\geq 160$ GeV and $p_T^{j_2,j_3,j_4} \geq 60$ GeV. For all other jets we demand $p_T^j \geq 20$ GeV. The rapidity coverage of the jets is determined by ATLAS calorimeter design, where the forward calorimeter covers the pseudo-rapidity range of $|\eta|<4.9$. However, the tracker covers only upto $|\eta|<2.5$, and therefore it is not possible to obtain the information on the number of charged tracks inside jets in the forward region. Since the quark-gluon discrimination variables can be more accurately determined in terms of charged track momenta, we therefore count $n_j$ only within $|\eta|<2.5$. 

\item No isolated lepton (electron or muon) with $p_T>10$ GeV, within $|\eta|<2.5$. 

\item Missing transverse momentum in the event $\met > 160$ GeV.

\item $\Delta \phi { ({\rm jet}, \met)_{\rm min}} > 0.4$ (0.2) radian for $j_1, j_2, j_3$ (for all other jets with $p_T>40$ GeV).
\end{enumerate}

The jet $p_T$ cuts and the $\met$ cut are applied at the matrix element (ME) level while generating the background events, which is modelled by $Z(\rightarrow\nu \bar{\nu}+$)jets. Furthermore, in order to obtain a large statistics of events with a high $M_{\rm eff}$ cut, we have generated several different samples of the $Z+$jets events, one with each value of the $M_{\rm eff}$ cut. As discussed in detail in Sec.~\ref{MC}, we normalize our total $Z+$jets event rate by comparison with the number of events reported in the ATLAS simulation after the cuts in {\tt 4jm} category~\cite{ATLAS14} (defined as {\tt Cut-1} followed by $\met /M_{\rm eff} > 0.25$ and $M_{\rm eff}>3200$ GeV). With this, we are able to reproduce with a reasonable accuracy the ATLAS projected sensitivity in the $M_{\gl}-M_{\neu}$ plane for $14$ TeV LHC with $300 {~\rm fb}^{-1}$ data.

In addition to the above basic set of cuts, in order to compare with the search reach of ATLAS 14 TeV projections~\cite{ATLAS14}, we have computed the signal and background event yields in seven different signal regions ({\tt 4jl, 4jm, 4jt, 5j, 6jl, 6jm, 6jt}) as defined in the ATLAS study~\cite{ATLAS14}, essentially differing in the values of the $M_{\rm eff}$, $\met /M_{\rm eff}$ and $\met/\sqrt{H_T}$ cuts.

\subsection{Monte Carlo simulation of signal and background processes}
\label{MC}
For both the signal and background processes, the parton level matrix elements are computed, and the events generated using {\tt MG5aMC@NLO}~\cite{MG5}. The parton level events are passed onto both {\tt Pythia 6.4.28} (with the {\tt P2012-RadLo} tune)~\cite{Pythia}, and {\tt Herwig++ 2.7.1} (with the default tune)~\cite{Herwig}, for simulating parton shower, hadronization and underlying events. The above choice for the {\tt Pythia} tune is based on better data-model agreement in a recent ATLAS study comparing the charged track multiplicity distribution in the data with MC predictions~\cite{ATLAS_charged}. The parton shower and hadronization effects are simulated using two different MCs to estimate the uncertainty in quark-gluon tagging coming from MC modelling. The signal cross-section is normalized to predictions including the resummation of soft-gluon emission at next-to-leading logarithmic accuracy, matched to next-to-leading order supersymmetric QCD corrections
~\cite{SUSY_Working}. 

We use the {\tt CTEQ6L1}~\cite{Cteq} parton distribution functions from the {\tt LHAPDF}~\cite{LHAPDF} library, and the factorization and renormalization scales are kept at the default event-by-event choice of {\tt MG5aMC@NLO}. Detector effects have been simulated using {\tt Delphes3}~\cite{Delphes}, where the jet clustering is performed with {\tt FastJet3}~\cite{Fastjet}. Jets are reconstructed using the anti-$k_T$ clustering algorithm~\cite{Fastjet,antikt} with radius parameter $R=0.4$. We have implemented the variables used for studying quark and gluon jet tagging in the {\tt Delphes3} framework. 

As the signal process, we consider gluino pair production, followed by its three-body decay with $100\%$ branching ratio to a pair of quarks and the lightest neutralino, via intermediate off-shell squarks. In general, depending on the squark mass, on-shell squark production will also contribute to the same final state. However, for studying the usefulness of quark-gluon tagging in SUSY searches, a simplified model with only the gluino and the lightest (bino-like) neutralino is adequate, and the rest of the MSSM particles are assumed to be decoupled. The final state of interest will then be $\geq 4$-jets and missing transverse momentum.

It is well-understood that the primary background to such a multi-jet and missing momentum search comes from $Z+$jets production (with $Z$ decaying to neutrinos), followed by a similar contribution from $W+$jets (where the charged lepton from the $W$ boson decay falls outside the tracker acceptance, and therefore is not reconstructed as a lepton). The fractional contribution of $t\bar{t}+$jets and single top production is reduced at higher $M_{\rm eff}$ regions, but it can also become comparable to the individual weak-boson contributions depending upon the signal region of interest. A strong cut on the $\met$ variable reduces the QCD multijet background, especially by ensuring that the jet direction and the $\met$ vector direction are not correlated. For a comparison of different SM background contributions, see, for example, the recent ATLAS note on squark-gluino search at the 13 TeV LHC with $13.3 {~\rm fb}^{-1}$ of data~\cite{ATLAS_13TeV_ICHEP}. Both the recent $13$ TeV ATLAS analysis and the ATLAS projection results for $14$ TeV LHC with $300 {~\rm fb}^{-1}$ data show that the total SM background in our signal region of interest (i.e., after {\tt Cut-1} and with $M_{\rm eff}>1.8$ TeV) is always less than twice the $Z+$jets contribution. The kinematic and quark-gluon fraction properties in $Z+$jets and the subdominant $W+$jets processes are nearly identical. Therefore, we perform the MC simulations using only the $Z+$jets process, and take the total SM background as twice the $Z+$jets prediction, which is a conservative estimate.

Since we shall focus on a multivariate analysis (MVA) strategy especially for the quark-gluon separation, the statistics of MC events required  to perform the boosted decision tree (BDT) training is very high, especially if the number of input variables to the BDT training is large (eventually we shall use a ten variable BDT). Furthermore, these event samples are all required to pass a pre-selection of {\tt Cut-1} and different values of high $M_{\rm eff}$ cuts. Therefore, generating such a large statistics of events with matrix element (ME) - parton shower (PS) matching is beyond the scope of our computational resources. On the other hand, as is well-known, to obtain accurate predictions for the jet $p_T$s in processes such as $Z+$jets, ME-PS matching is important. However, since we are primarily interested in four relatively hard and central jets, the expectation is that events based on $Z+3-$jets or $Z+4-$jets matrix elements followed by PS can cover the relevant phase space region, and therefore the normalized differential distributions should be well-predicted by these event samples. In order to check this fact, we generated three different samples of $Z+$jets events and compared all the kinematic and jet-substructure distributions between them. The three samples are: (1) $Z+$jets, ME-PS merged upto $4-$jets, (2) $Z+3-$jet ME followed by PS and (3) $Z+4-$jet ME followed by PS. We find that all the distributions have very similar shape in the three samples (as shown in the Appendix). Thus it is possible to obtain accurate normalized distributions by just using the $Z+3-$jet ME (followed by PS) event sample, for which generating a large enough statistics is least resource intensive among the three. For the overall normalization, as discussed earlier, we normalize our $Z+$jets event yield to the number reported in ATLAS simulation~\cite{ATLAS14}, and take the total SM background as two times the $Z+$jets contribution. 

\section{Results}
\label{results}
\subsection{MC truth level quark-gluon fraction}
\label{sec:MC_truth}
As discussed in Sec.~\ref{MC}, in the signal process of gluino pair production, with gluino dominantly decaying via (onshell or offshell) squarks, the decay jets are all quark-initiated. In addition, there are additional jets in the signal events from initial state radiation (ISR), which may reduce the difference between the signal and background likelihoods if a gluon-initiated ISR jet is harder than the decay jets and also lies in the central region of the detector. At Born level, the dominant background of $Z+$jets has a higher gluon fraction in the third and fourth highest $p_T$ jets (denoted by $j_3$ and $j_4$ respectively). It is thus expected that the maximum discriminating power in the likelihood would come from $j_3$ and $j_4$, rather than the first and second highest $p_T$ jets (denoted by $j_1$ and $j_2$). 

To define the MC truth level quark and gluon jet fraction, we adopt the following method. Assume that we are looking for quark jets in an event. In the first step we find quarks in the matrix element, and a quark of flavour $f$ is denoted by $f_i$. Next, in the parton history related to the mother parton $i$, we find the parton $P_i$ with the same flavour as $f_i$ (we choose the parton with the highest transverse momentum if there are multiple quark partons of flavour $f$). Finally, if the distance between the jet $J$ and the parton $P_i$ is less than the jet cone size, $\Delta R (J,P_i)< R = 0.4$, we define the jet $J$ as a quark jet. If not, then $J$ is defined as a gluon jet. We emphasize that in the actual study of signal-background discrimination, this definition does not play any role, since in that case, we compare the likelihood of an event being signal-like or background-like, based on an MVA with the discriminating variables as inputs.

For illustration, we show in Tab.~\ref{tab:qg} the parton level quark fraction of the first four jets, as defined above. A representative signal point with $M_{\gl}=2000$ GeV and $M_{\neu}=1000$ GeV has been chosen for Tab.~\ref{tab:qg}, and the quark fractions are shown after the preselection of {\tt Cut-1} and with $M_{\rm eff} > 1.8$ TeV. The parton shower MC used for this figure is {\tt Pythia 6.4.28}. In general, we see from this table that among the first four hardest jets, most signal events contain $3-4$ quark jets, while most $Z+$jets events contain $1-2$ quark jets. 
\begin{table}[htb!]
\centering
\begin{tabular}{|c|c|c|c|c|}
\hline
Process & $f_q^{j_1}$ & $f_q^{j_2}$ &  $f_q^{j_3}$ & $f_q^{j_4}$  \\
\hline
$\gl \gl +$jets& 0.92       & 0.87 &     0.77             &  0.64 \\

$Z+$jets &   0.64           &0.55 &     0.27              & 0.16  \\
\hline
\end{tabular}
\caption{\small \sl Quark fraction ($f_q$) at the MC truth level for the first four highest-$p_T$ jets in $\gl \gl +$jets and $Z+$jets processes. All events are selected after passing the jet-$p_T$, $\met$ ({\tt Cut-1}) and $M_{\rm eff}>1.8$ TeV cuts, at the 14 TeV LHC. See text for details on the determination of $f_q$.}
\label{tab:qg}
\end{table}

\subsection{Inclusive and exclusive kinematic variables}
\label{sec:kin_excl}
\begin{figure}[htb!]
\centering 
\includegraphics[width=0.45\textwidth]{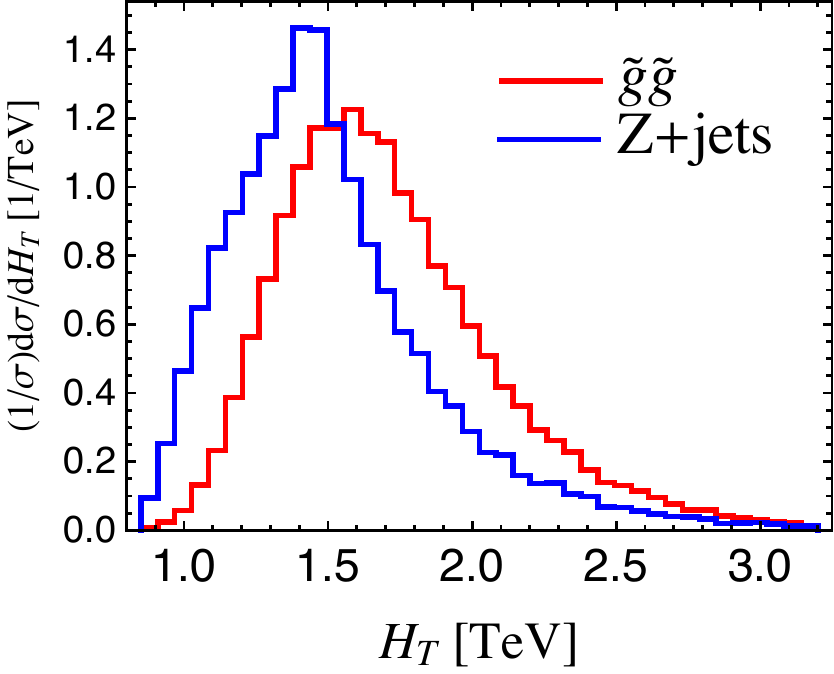}
\hfill 
\includegraphics[width=0.45\textwidth]{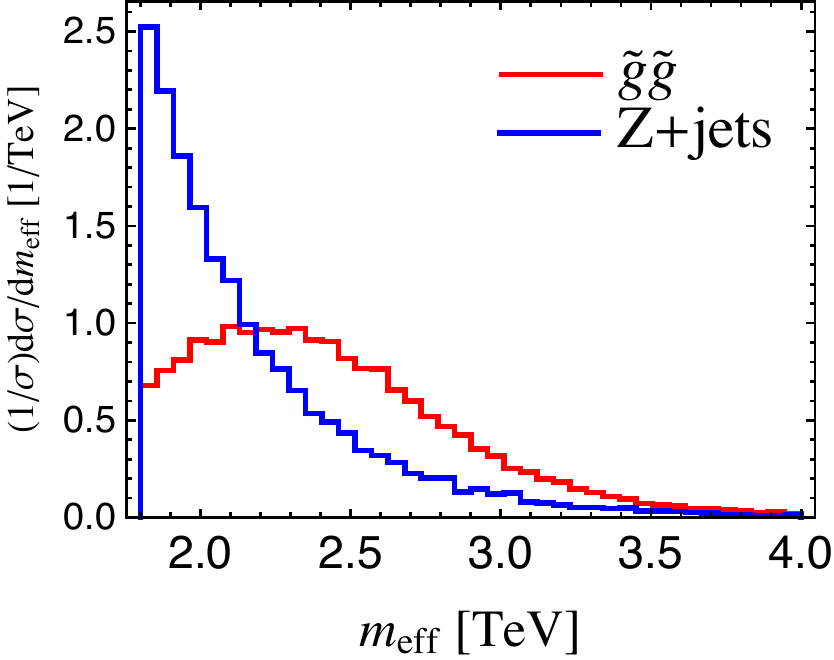}
\hfill
\includegraphics[width=0.45\textwidth]{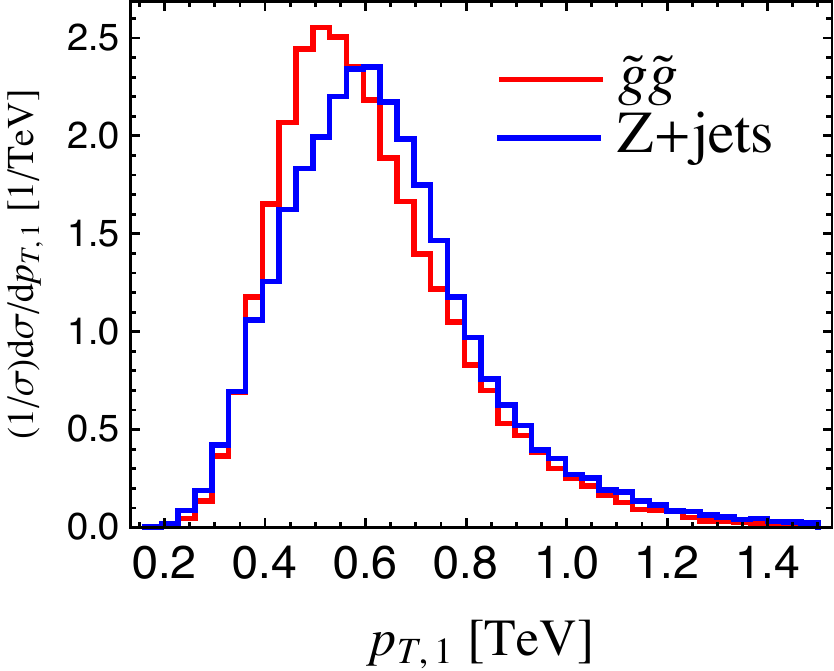}
\hfill
\includegraphics[width=0.45\textwidth]{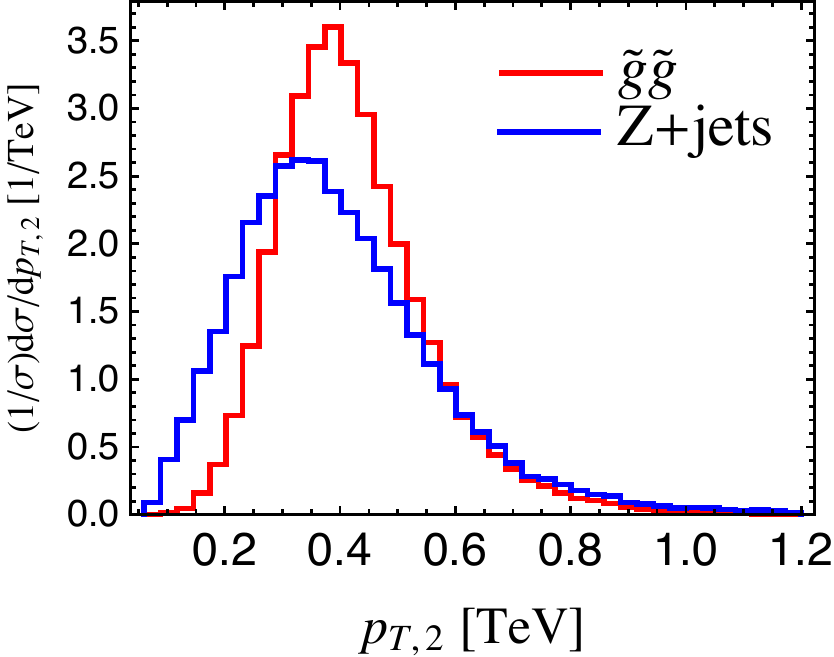}
\hfill
\includegraphics[width=0.45\textwidth]{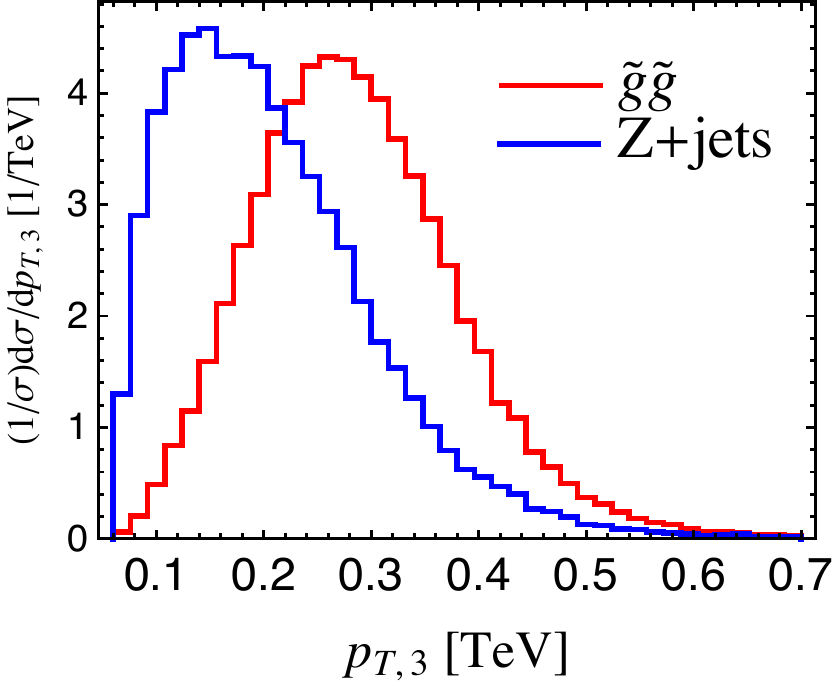}
\hfill
\includegraphics[width=0.45\textwidth]{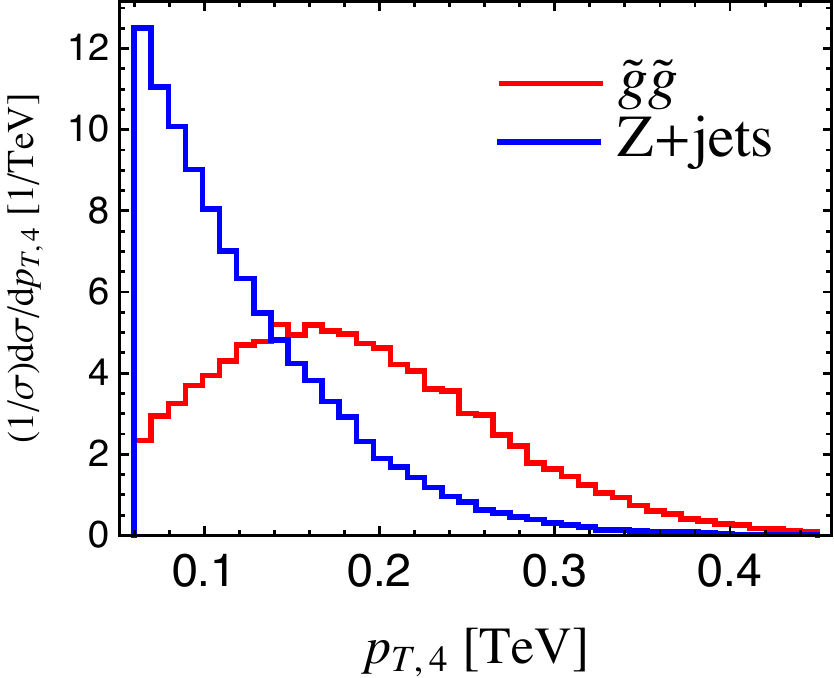}
\caption{\label{fig:kin}\small \sl Normalized distribution of inclusive and exclusive kinematic variables. For the signal, we show the distributions at a benchmark point with $M_{\gl} = 2000$ GeV and $M_{\neu}=1000$ GeV. The distributions are presented after {\tt Cut-1} and an $M_{\rm eff}$ cut of $1.8$ TeV, and for this reason the $H_T$ and $p_{Tj}$ distributions have a non-standard shape.}
\end{figure}

In the dominant SM background processes of $Z/W+$jets, the jets come from initial state QCD radiation, which exhibits a strong ordering of the jet $p_T$s for a given $H_T$ value, primarily because of the enhancement in the soft gluon emission probability given by the QCD splitting functions. On the other hand, for the decay jets coming from gluino decay, the jet transverse momenta are not in general so strongly ordered, as in this case the $p_T$s of the jets are determined by the mass-gap between the gluino and the lightest neutralino and the mass of the lightest neutralino itself. Admittedly, this is then a SUSY parameter dependent statement as to how the ordered jet $p_T$ distributions would differ between the signal and the background. Nevertheless,  for certain ranges of the gluino and neutralino masses, the transverse momentum of the first four jets, ordered according to their $p_T$s, can carry additional information not entirely captured in the $M_{\rm eff}$ or $H_T$ distributions. We use the nomenclature of exclusive kinematic variables to refer to the ordered jet $p_T$s, while we shall refer to $M_{\rm eff}, H_T$ and $\met$ as inclusive kinematic variables.  

We show in Fig.~\ref{fig:kin}, the normalized (to unit area) distributions for the kinematic variables used as inputs in defining the combined signal and background likelihood functions, after the event pre-selection of {\tt Cut-1} and an $M_{\rm eff}$ cut of $1.8$ TeV. For the signal, we show the distributions at a benchmark point with $M_{\gl} = 2000$ GeV and $M_{\neu}=1000$ GeV, and for illustration results from only the {\tt Pythia} MC are presented. Since only events passing {\tt Cut-1} and $M_{\rm eff}>1.8$ TeV are included, the $H_T$ and $p_{Tj}$ distributions have a non-standard shape (first rise to a peak value and then fall). As we can see from this figure, for this signal benchmark point, the exclusive kinematic variables also provide discriminating power over the $Z+$jets background. For the gluino pair production events, we have also checked that including additional jets in the matrix element and using ME-PS matching, the kinematic distributions do not show any significant difference.

\subsection{Multivariate analysis}
\label{sec:qg_BDT}
\begin{figure}[htb!]
\centering 
\includegraphics[width=0.45\textwidth]{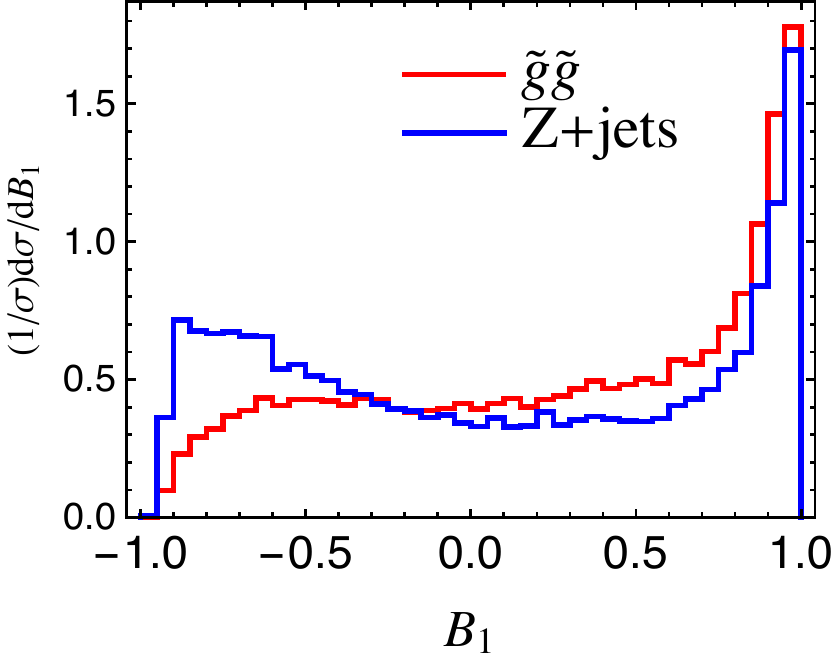}
\hfill 
\includegraphics[width=0.45\textwidth]{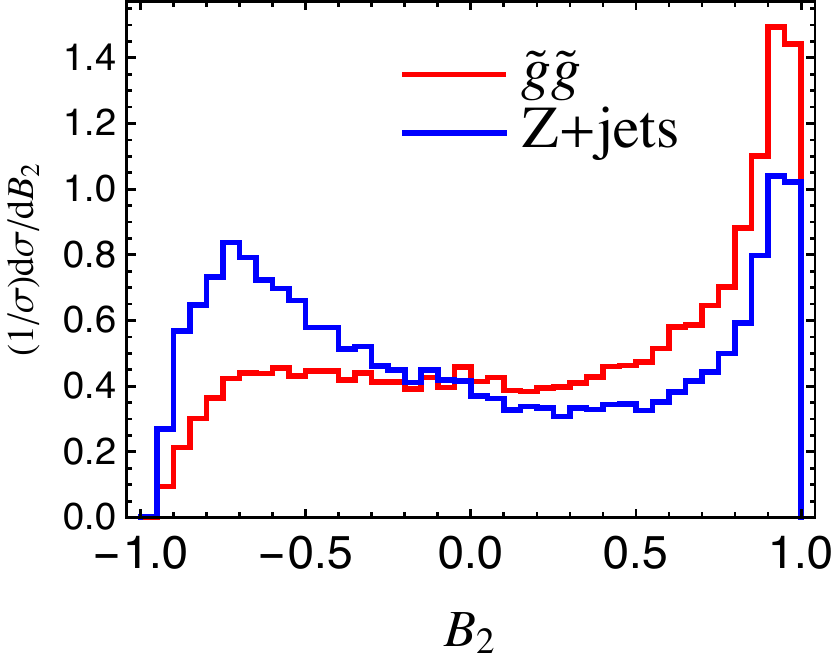}
\hfill
\includegraphics[width=0.45\textwidth]{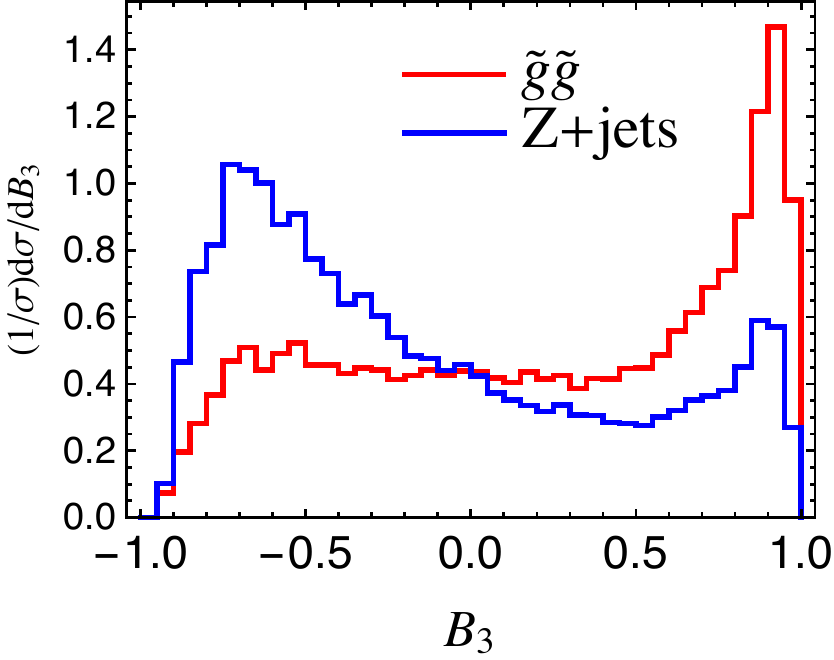}
\hfill
\includegraphics[width=0.45\textwidth]{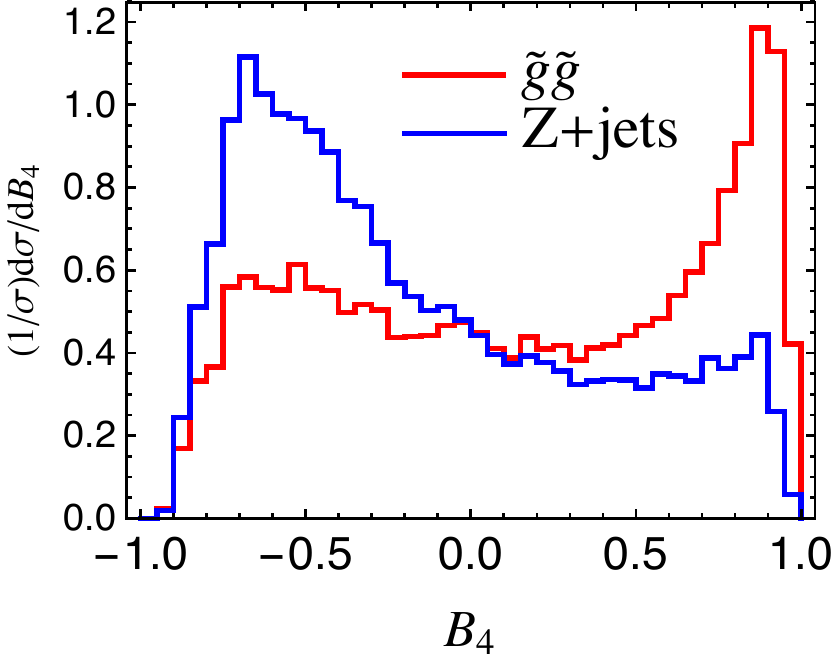}

\caption{\label{fig:BDT_qg}\small \sl Normalized BDT score distributions based on quark-gluon tagging variables of individual jets ordered according to their $p_T$ ($B_i$ refers to the BDT score for $j_i$). For illustration, the distributions are shown for a signal benchmark point of $M_{\gl} = 2000$ GeV and $M_{\neu}=1000$ GeV, and using the {\tt Pythia6} MC.}
\end{figure}
Using the quark-gluon separation variables described in Sec.~\ref{sec:qg_vari}, namely, $n_{\rm ch}, C_1^{(\beta)}$ and $m_J/p_{T,J}$ as inputs, we first develop an optimized discriminant using a multivariate analysis. This has been carried out by employing a Boosted Decision Tree (BDT) algorithm with the help of the {\tt TMVA-Toolkit}~\cite{TMVA} in the {\tt ROOT} framework~\cite{ROOT}. The training of the BDT classifier has been performed using the $Z+q$ and $Z+g$ processes at the Born level. The MC samples for these processes are generated such that we obtain an uniform statistical coverage across the entire jet $p_T$ range of interest, and the BDT training is performed for different $p_T$ ranges taken as different categories. 

Following the above method, for the signal and background processes, we compute the BDT score $B_i$ for each of the first four jets ordered according to their $p_T$. This procedure has been carried out using both the {\tt Pythia6} and {\tt Herwig++} MCs to simulate the parton shower and hadronization aspects. In Fig.~\ref{fig:BDT_qg}, we show the distribution of the BDT scores for the first four highest $p_T$ jets in the gluino pair signal and the $Z+$jets background processes (for illustration, the distributions are shown using {\tt Pythia6}). As expected from the truth level quark-gluon fractions discussed in Sec.~\ref{sec:MC_truth}, significant separation in the BDT scores for the third and fourth highest $p_T$ jets ($B_3$ and $B_4$) are observed, for which the signal jets are mostly quark-initiated, and the background ones are mostly gluon-initiated.

\begin{figure}[htb!]
\centering 
\includegraphics[width=0.45\textwidth]{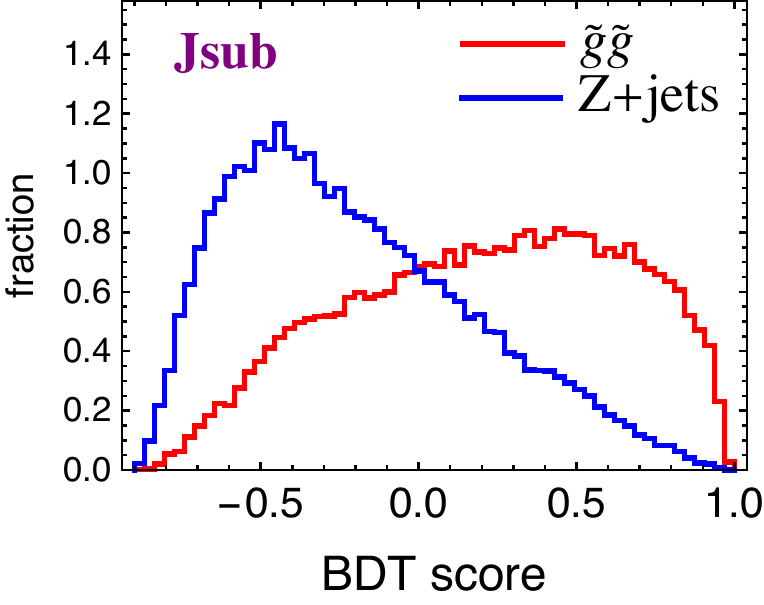}
\hfill 
\includegraphics[width=0.45\textwidth]{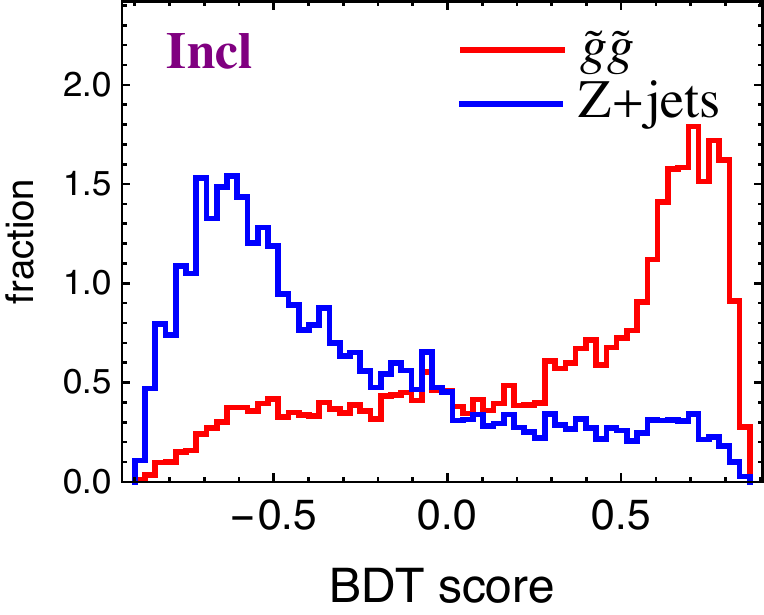}
\hfill
\includegraphics[width=0.45\textwidth]{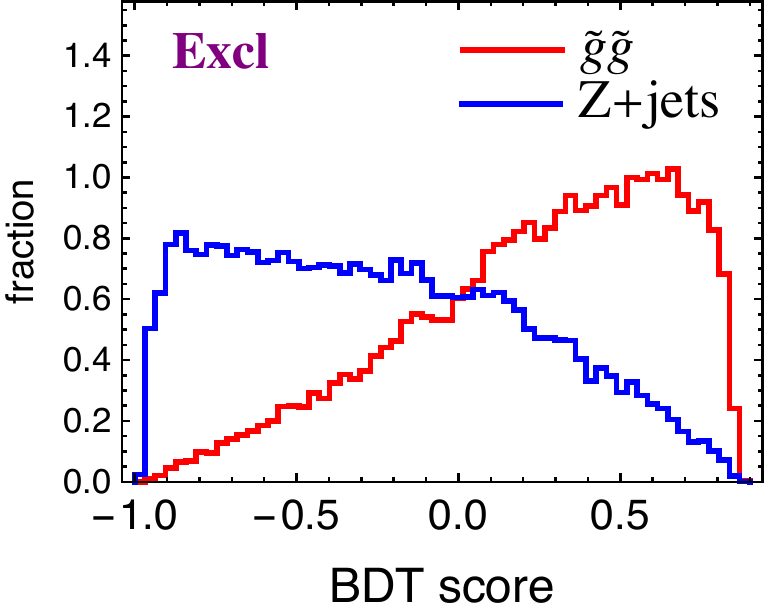}
\hfill
\includegraphics[width=0.45\textwidth]{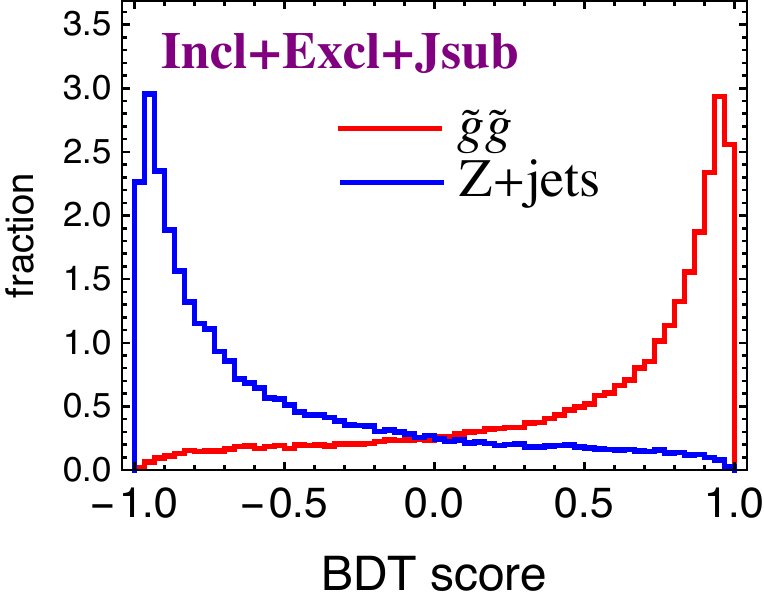}
\caption{\label{fig:BDT_sb}\small \sl Signal and background likelihood distributions, using as inputs different subsets of variables. {\tt Incl.}, {\tt Excl.} and {\tt Jsub} refer to MVA with the input sets $\{M_{\rm eff}, H_T\}$, $\{p_{T,j1}, p_{T,j2}, p_{T,j3}, p_{T,j4}\}$, and $\{B_1, B_2, B_3, B_4\}$ respectively. The bottom right plot is obtained using an MVA with all ten variables as inputs.}
\end{figure}

As a final ingredient to our analysis, we perform a further MVA study with ten input variables containing:  \{$M_{\rm eff}, H_T, p_{T,j1}, p_{T,j2}, p_{T,j3}, p_{T,j4}, B_1, B_2, B_3, B_4$\}. This defines a signal and background likelihood with all the kinematic and jet substructure information of the event. The BDT score cut is chosen to maximize the exclusion (or discovery) significance for a given model point. For illustrating the separation power from each subset of variables, we show in Fig.~\ref{fig:BDT_sb} the BDT score distributions obtained with the inclusive kinematic variables ($M_{\rm eff}$ and $H_T$), the exclusive kinematic variables ($p_{T,j1}, p_{T,j2}, p_{T,j3}$ and $p_{T,j4}$), and the jet substructure based BDT variables ($B_1, B_2, B_3$ and $B_4$). We also show in the bottom right panel of Fig.~\ref{fig:BDT_sb} the signal-background separation with all ten variables included together in the MVA. The results are shown for the signal point $(M_{\gl}=2000 {~\rm GeV}$ and $M_{\neu}=1000 {~\rm GeV})$ and with {\tt Pythia} MC.
\begin{figure}[htb!]
\centering 
\includegraphics[width=0.5\textwidth]{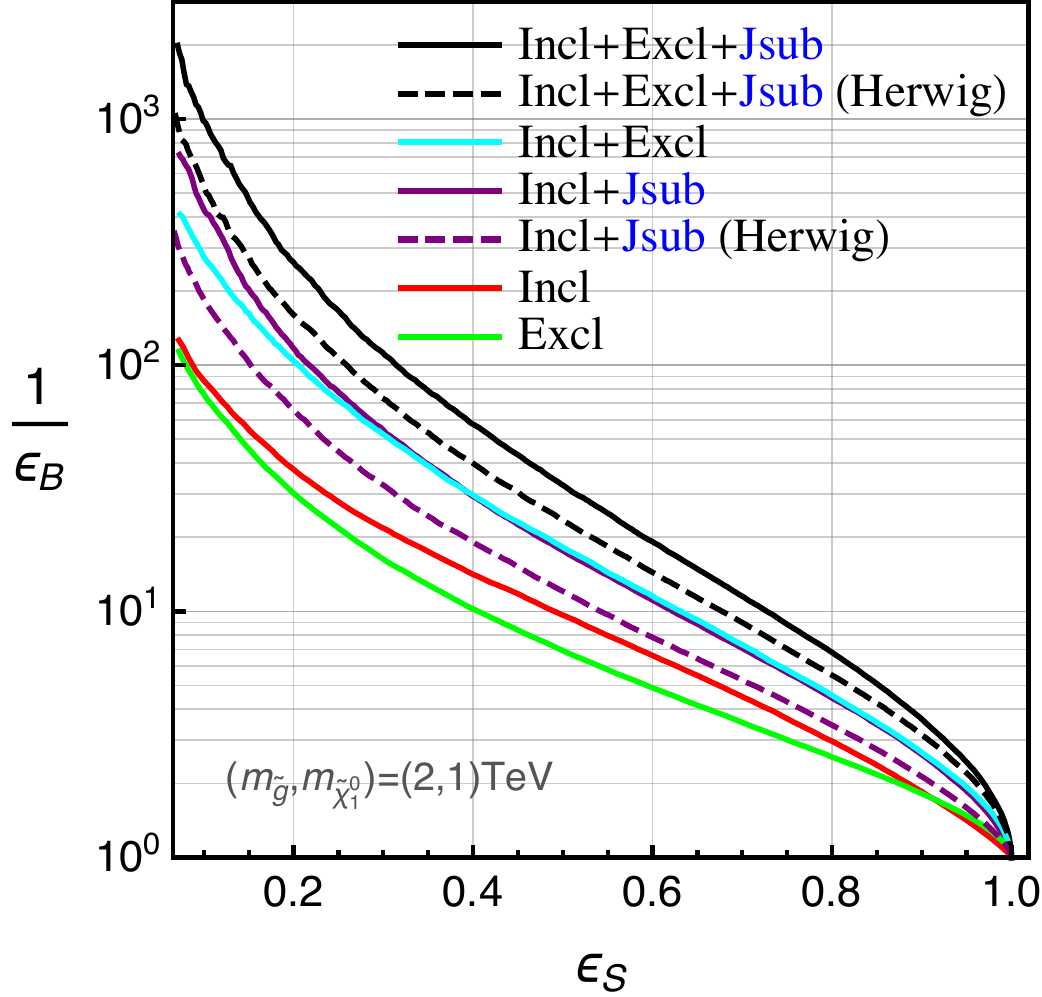}
\caption{\label{fig:roc}\small \sl Signal acceptance ($\epsilon_S$) versus inverse of the background acceptance ($1/\epsilon_B$)  efficiencies as a function of the BDT cut, for different MVA methods. The solid lines are obtained using {\tt Pythia6}, while the dashed ROC curve is obtained using the {\tt Herwig++} MC. The ROC curves predicted by {\tt Herwig++} in the {\tt Incl}, {\tt Excl} and {\tt Incl+Excl} categories are similar to the {\tt Pythia6} ones. See text for details.}
\end{figure}

Based on the final BDT score distribution with ten observables, we can now obtain the ROC curve which shows the signal acceptance ($\epsilon_S$) versus background rejection ($1-\epsilon_B$) efficiencies as a function of the BDT cut. In Fig.~\ref{fig:roc}, the red, green and cyan curves show the ROC curves for the MVA analyses based on the inclusive, exclusive and inclusive and exclusive variable sets combined respectively. These two sets carry independent information, and therefore the background rejection using the combined set increases compared to the ones using the individual sub-sets, and the $\epsilon_S$ and $1/\epsilon_B$ values on the cyan curve is roughly given by the product of their corresponding values on the red and green curves. For example, the efficiencies ($\epsilon_S$, $\epsilon_B^{-1}$) for the red and the green curves pass through $(0.4, 10)$ and $(0.4, 1.4 \times 10)$ respectively, and the cyan curve passes through the product $(0.4^2, 1.4 \times 10^{2})$. The black solid and dashed curves show the performance of the MVA analysis with all the variables taken together, using {\tt Pythia6} and {\tt Herwig++} respectively.  We find that, by adding the jet substructure variables to the MVA, the background rejection factor increases by about a factor of $4$ in the {\tt Pythia} results and by a factor of $2-2.5$ in the {\tt Herwig} results, for $\epsilon_S \sim 0.1$. As we shall see in the next subsection, this latter improvement has a considerable impact while considering the exclusion (discovery) reach in the $M_{\gl}-M_{\neu}$ plane. For comparison of different combination of variables, we also show the ROC curves with the combination of inclusive and jet substructure observables, using the violet solid ({\tt Pythia}) and violet dashed ({\tt Herwig}) curves.

\subsection{Projected reach in $M_{\gl}-M_{\neu}$ plane}
By varying the BDT score cut with all or a subset of observables as input, we can choose the cut that maximizes the search significance for a given SUSY parameter point. Here, we use the profile-likelihood method~\cite{PLmethod} to determine the $95\%$ C.L. exclusion region in the $M_{\gl}-M_{\neu}$ plane. The likelihood function is defined as follows:
\begin{equation}
\mathcal{L}(N_{\rm obs}|B+S) \propto \max_{B^\prime=\{0,\infty\}}\frac{e^{-(S + B^\prime)}\left(S + B^\prime \right)^{N_{\rm obs}}}{N_{\rm obs} !} \exp \left[ -\frac{(B^\prime-B)^2}{2\sigma_B^2}\right]~, 
\label{LHC-like}
\end{equation}
where $S$ and $B$ denote the expected number of signal (for a given point in the parameter space) and background event yields with a particular integrated luminosity, and 
$N_{\rm obs}$ represents the number of events observed in the corresponding search with the same luminosity. For determining the exclusion contours, we set $N_{\rm obs}$ to be equal to the mean value of the expected number of background events $B$.

The systematic uncertainty in the background prediction is taken into account by convoluting the Poission likelihood function with a Gaussian with mean $B$ and variance $\sigma_B$. Combining the different components of the systematic uncertainty, we set $\sigma_B=(\delta_{\rm Incl}+\delta_{\rm Excl}+\delta_{\rm Jsub}) \times B$, where $\delta_{\rm Incl},\delta_{\rm Excl}$ and $\delta_{\rm Jsub}$ are the fractional systematic uncertainties in the background prediction coming from inclusive, exclusive and jet substructure observables respectively. For our significance computation we have set $\delta_{\rm Incl}=\delta_{\rm Excl}=\delta_{\rm Jsub}=10\%$, and to obtain a conservative estimate of the reach we have added these uncertainties linearly, making the total systematic uncertainty in the background yield prediction to be $30\%$, when all the variables are included together. We introduced a nuisance parameter $B^\prime$ to deal with the systematic uncertainty, which is profiled out by maximizing the likelihood function by varying $B^\prime$ in the interval $0 \leq B^\prime \leq \infty$.

\begin{figure}[htb!]
\centering 
\includegraphics[width=0.6\textwidth]{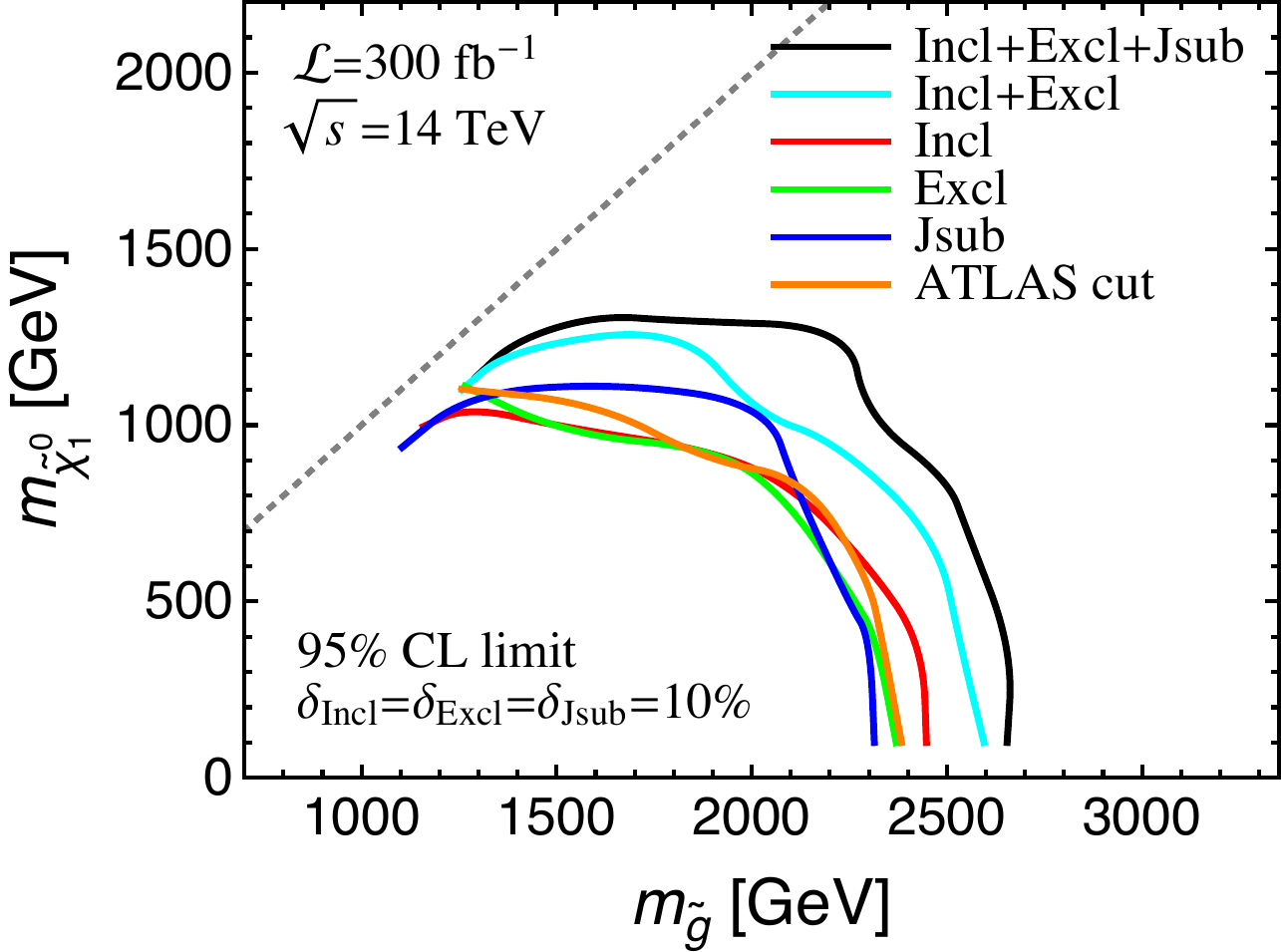}
\caption{\label{fig:reach}\small \sl Projected $95\%$ C.L. exclusion contours in the $M_{\gl}-M_{\neu}$ plane at the 14 TeV LHC with an integrated luminosity of $300 {~\rm fb}^{-1}$. The different systematic uncertainty components have been added linearly, making the total systematic uncertainty in the background yield prediction to be $30\%$, when all the variables are included together. See text for details on the individual exclusion contours.}
\end{figure}

In Fig.~\ref{fig:reach}, we show the projected $95\%$ C.L. exclusion contours in the $M_{\gl}-M_{\neu}$ plane at the 14 TeV LHC with an integrated luminosity of $300 {~\rm fb}^{-1}$. The orange curve is the ATLAS projected sensitivity with standard kinematic cuts (as reproduced by us), while the red, green and blue solid lines show the reach with each subset of variables described in the previous subsection. As expected from the ROC curve in Fig.~\ref{fig:roc}, each of the subsets individually can lead to similar reach in this parameter space. We recall that all the curves include the effect of the pre-selection cuts on the jet $p_T$'s and $\met$ ({\tt Cut-1}) as well as a high $M_{\rm eff}$ cut. Thus these improvements are within a high mass signal region. It is further observed that on including the information of the ordered jet $p_T$s of the first four jets the reach improves to a good extent (cyan solid curve). Finally, if we now include the jet substructure information as well, the reach in the $M_{\gl}-M_{\neu}$ plane (black solid line) shows considerable improvement over the standard analysis. It should be noted in particular that especially in the region where the mass difference between the gluino and the neutralino falls in an intermediate range, the jet substructure observables provide stronger separation power. We also note that the signal benchmark point used to show the various distributions in this study, namely, $(M_{\gl}=2000 {~\rm GeV}, M_{\neu}=1000 {~\rm GeV})$ can be excluded at $2\sigma$ level only when the jet substructure variables are included in the MVA. Since we have also included additional systematic uncertainties in the background rate coming from the modelling of both the exclusive and jet substructure observables (upto $30\%$ in total systematic uncertainty), our estimates for the improvement in the LHC reach should be conservative. It is thus promising that utilizing quark-gluon discrimination within an MVA including kinematic observables can considerably improve the LHC search prospects of strongly interacting SUSY particles.

\begin{figure}[htb!]
\centering 
\includegraphics[width=0.6\textwidth]{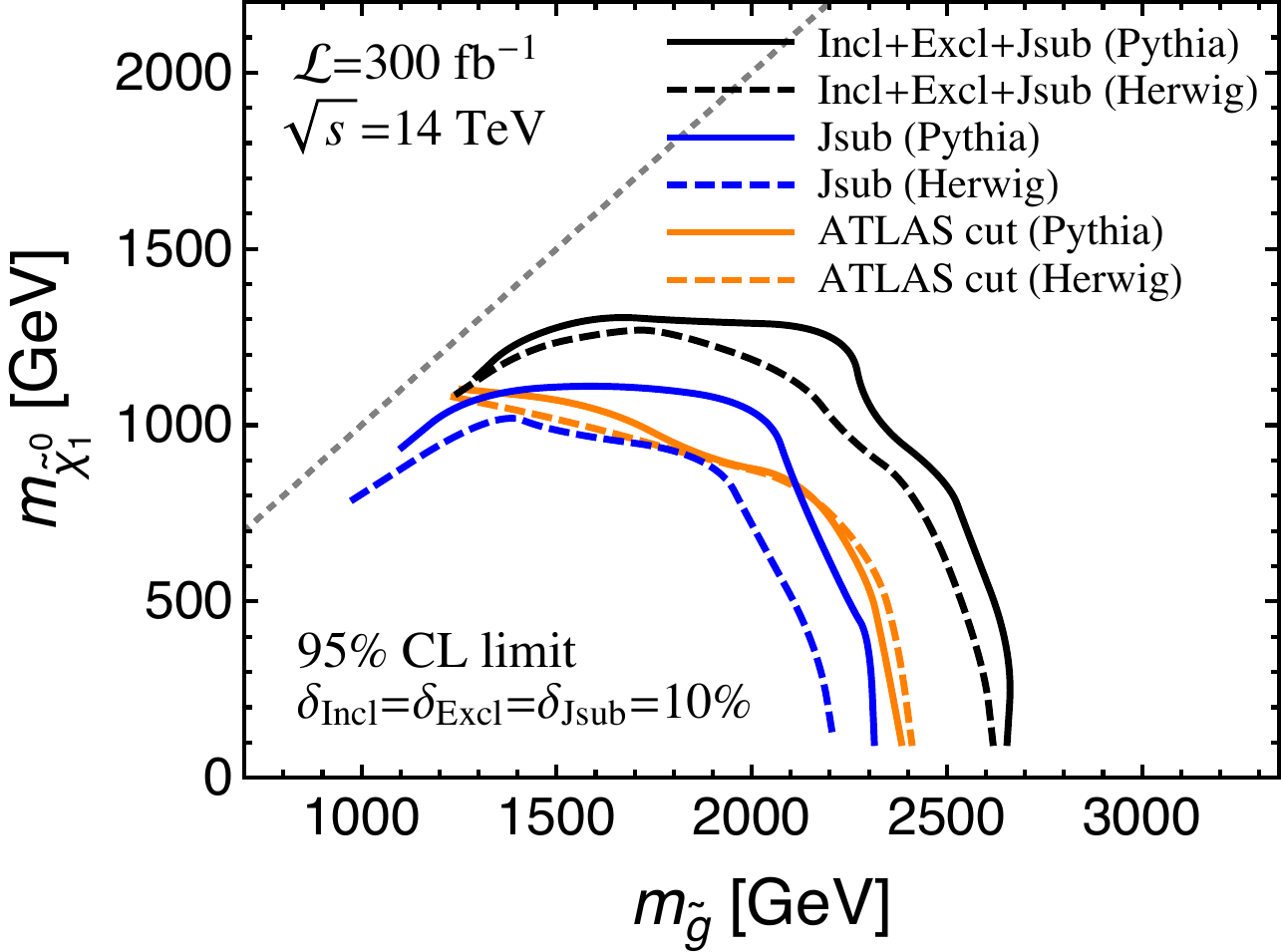}
\caption{\label{fig:py_vs_hw}\small \sl The $95\%$ C.L. exclusion contours predicted by {\tt Pythia6} (solid lines) and  {\tt Herwig++} (dashed lines) using either only the jet substructure subset (blue curves) or the full variable set (black curves). For reference, the exclusion contours based on ATLAS cuts~\cite{ATLAS14} are also shown (orange curves), and they are almost identical for {\tt Pythia6} and {\tt Herwig++}.}
\end{figure}

In order to understand the uncertainty in the predictions from the MC modelling of jet substructure, we have performed the full analysis using both the {\tt Pythia6} and {\tt Herwig++} MCs. In Fig.~\ref{fig:py_vs_hw} we show the $95\%$ C.L. exclusion contours predicted by the two MCs using either only the jet substructure subset (blue curves) or the full variable set (black curves). For reference, the exclusion contours based on ATLAS cuts~\cite{ATLAS14} are also shown (orange curves), and they are almost identical for {\tt Pythia6} and {\tt Herwig++}. The {\tt Pythia6} exclusion contours (solid lines) show a better reach than the {\tt Herwig++} ones (dashed lines), and the difference between the two essentially comes from the jet substructure modelling, which, as remarked earlier, differs significantly for gluon jets. It is however encouraging that both MCs predict significant improvement over the standard analysis. Thus to the extent these two MCs provide an estimate of the uncertainty in prediction, our results show that irrespective of such differences, an improvement is expected in the LHC reach of gluino pair production, especially in the intermediate mass gap region, when we include the quark-gluon separation information within the MVA analysis. Future availability of data-based templates and improved MC tunes are expected to lead to more reliable predictions and a reduction of the systematics in the application of quark-gluon discrimination.

\section{Summary and Outlook}
\label{summary}
Quark-gluon discrimination is becoming a topic of growing interest, both in the theoretical and Monte Carlo front with improved jet substructure based observables being designed to capture the detailed pattern of QCD radiation, and on the experimental front with the development of data-based templates for tagging observables as well as validation of existing MC tunes. It is thus an ideal juncture when the importance of quark-gluon jet separation methods in the search for physics beyond the standard model should be thoroughly explored. With this goal in mind, in this paper, we studied the impact of including quark- and gluon-initiated jet discrimination in the search for gluino pair production events at the LHC. As seen in Tab.~\ref{tab:qg}, when ordered according to their transverse momenta, the third and fourth jets are more likely to be quark-initiated for the signal process, while for the dominant background of $Z/W+$jets, they are more likely to be gluon-initiated. With the quark and gluon separation variables of the number of charged tracks, energy correlation functions $C_1^{\beta}$, and jet mass ($m_J/p_{T,J}$) as inputs to a multivariate analysis, we first develop a BDT-based quark-gluon discriminant across a large range of jet $p_T$ using the $Z+q$ and $Z+g$ processes as the training samples. In addition to the standard ``inclusive'' kinematic variables of $\met$, $H_T$ and $M_{\rm eff}$, we also observe that for a given $H_T$ value, there is a strong ordering of the jet $p_T$s for the $Z+$jets background process, while for the signal process the jets are not so strongly ordered. This is of course a parameter point dependent statement, as the gluino-neutralino mass splitting and the mass of the lightest neutralino determines the ordering of the $p_T$s of the decay jets. However, in certain regions in the $M_{\gl}-M_{\neu}$ plane the inclusion of these ``exclusive'' kinematic variables within an MVA can help in increasing the signal to background ratio ($S/B$). We have explored different combinations of the inclusive, exclusive and jet substructure observables as MVA input variables to understand the importance of each category, and find that all three sub-categories, when added individually to a set of  pre-selection cuts and a minimum effective mass cut (chosen according to the working point in the $(M_{\gl}, M_{\neu})$ plane), lead to a similar improvement in $S/B$. Consequently, compared to an optimized kinematic-category based search (as currently carried out by the ATLAS and CMS collaborations), inclusion of the quark-gluon discrimination variables improves the reach in the $M_{\gl}-M_{\neu}$ plane, especially in a region where the difference between $M_{\gl}$ and $M_{\neu}$ falls in an intermediate range. This is because for such intermediate mass gaps, the $H_T$ and $M_{\rm eff}$ distributions in the signal can become similar to the SM background ones. Given the fact that the jet substructure based variables, as well as the inclusive and exclusive kinematic distributions can bring in additional systematic uncertainties in the background rate determination, we have included a total systematic uncertainty of $30\%$ on our background event yield, which should be a reasonable estimate. 

As discussed in the introduction, there exist differences in the Monte Carlo prediction of the quark-gluon separation observables, and the data-based templates for gluon-initiated jets tend to lie in between the predictions of the {\tt Pythia} and {\tt Herwig} MCs, while for quark-initiated jets the data-based templates largely agree with the MCs. With this observation in view, we carry out our complete analysis using both the MC event generators, in order to get an understanding of the variation in signal and background rates from MC modelling of parton shower and hadronization processes. This translates into a variation in the expected reach in the $M_{\gl}-M_{\neu}$ plane as well. While the expected improvement in reach does depend upon the MC generator used, the generic patterns remain the same. The reach based on different sets of kinematic variables are similarly predicted by both the event generators, as largely expected, since the low energy hadronization component does not enter in the jet transverse momentum distributions, while the effect of parton shower variation is weaker if we focus on high-$p_T$ jets only. Therefore, the MC variation almost entirely originates from the modelling of the jet substructure. It is however encouraging that independent of the MC generator used, the inclusion of quark-gluon discrimination leads to an improvement in probing the gluino pair production process, especially in the intermediate mass-gap region. This fact, combined with the future prospect of obtaining data-driven multivariate templates that do not rely on the MC modelling of the hadronization component (and possible improvements in the MC tunes as well), makes the utilization of quark-gluon discrimination in new physics searches sufficiently promising. We therefore expect that it would be explored in further detail by the LHC experimental collaborations in the future search for strongly interacting supersymmetric particles.

\section*{Appendix}
\begin{figure}[htb!]
\centering 
\includegraphics[width=0.45\textwidth]{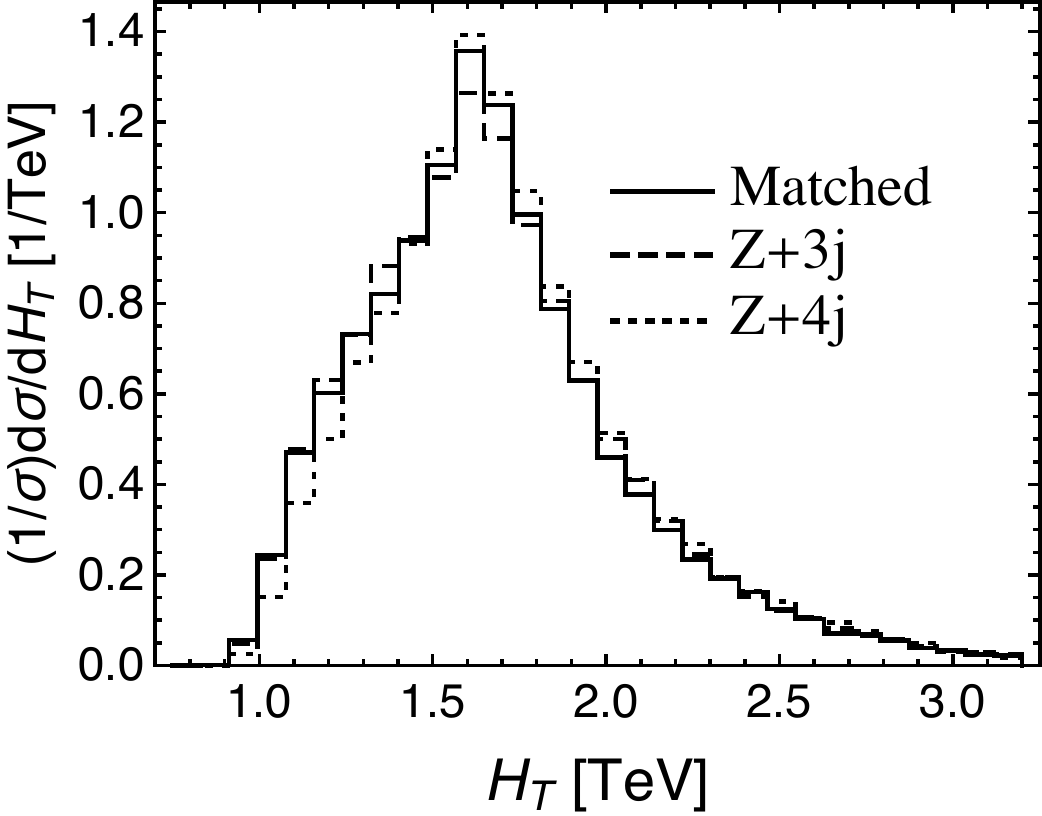}
\hfill 
\includegraphics[width=0.45\textwidth]{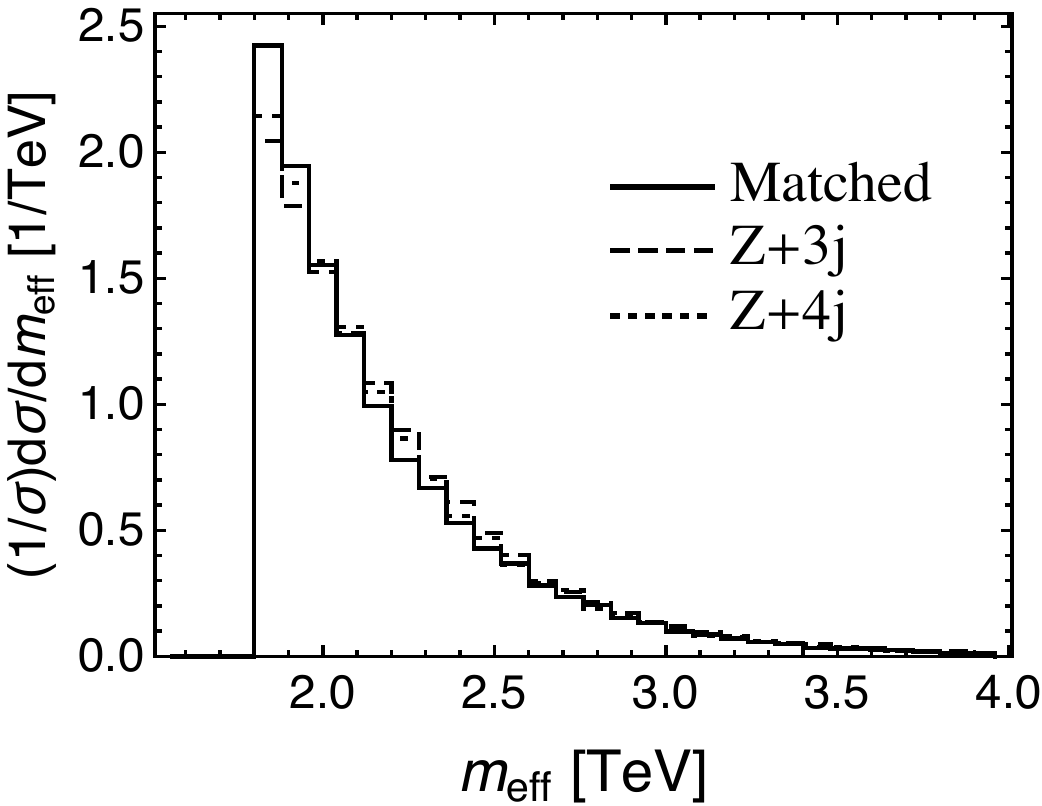}
\hfill
\includegraphics[width=0.45\textwidth]{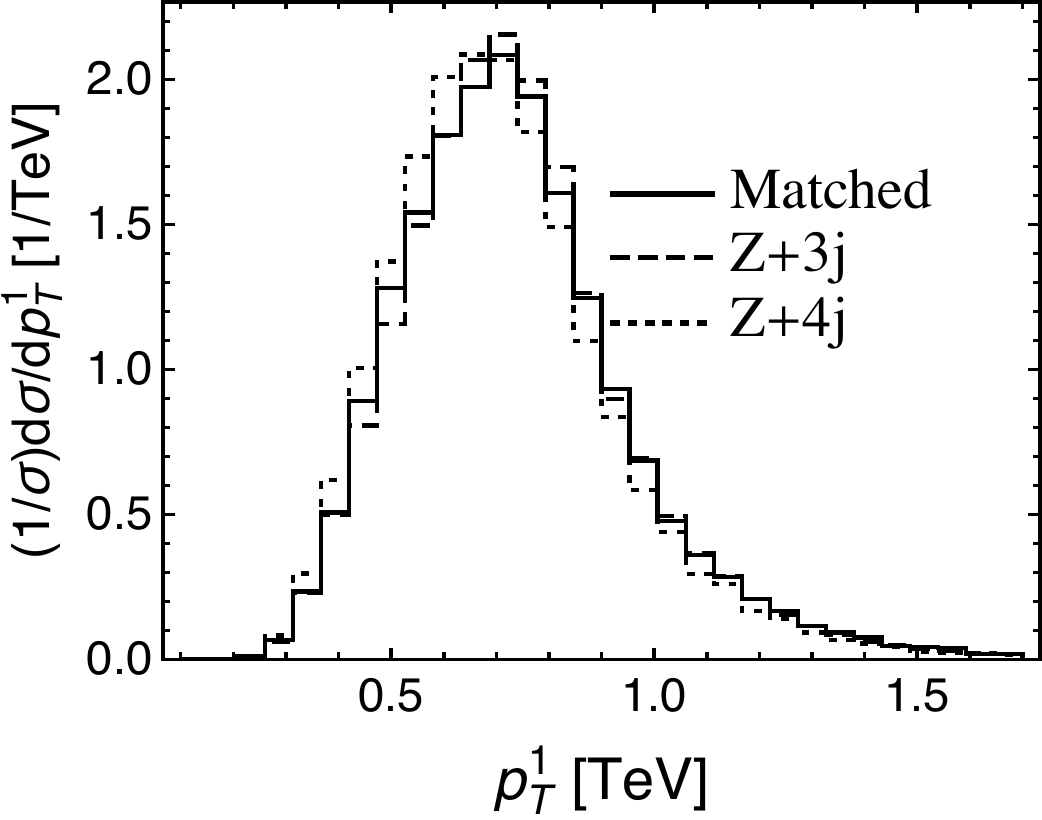}
\hfill
\includegraphics[width=0.45\textwidth]{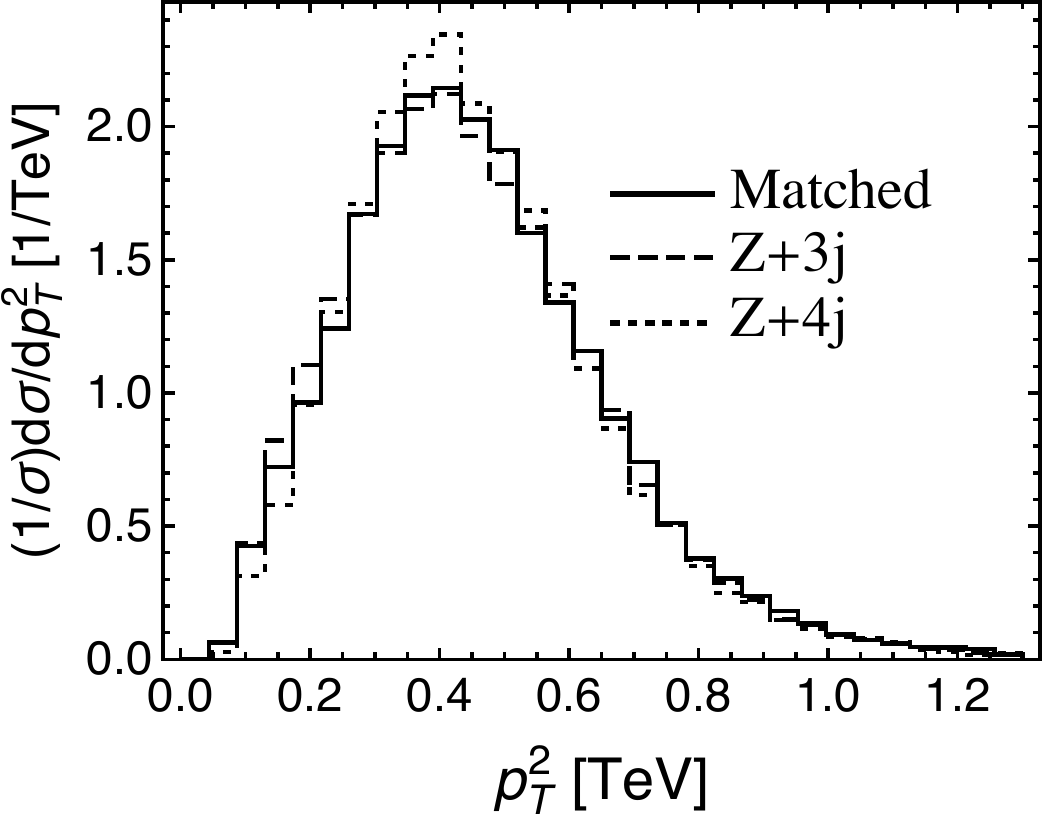}
\hfill
\includegraphics[width=0.45\textwidth]{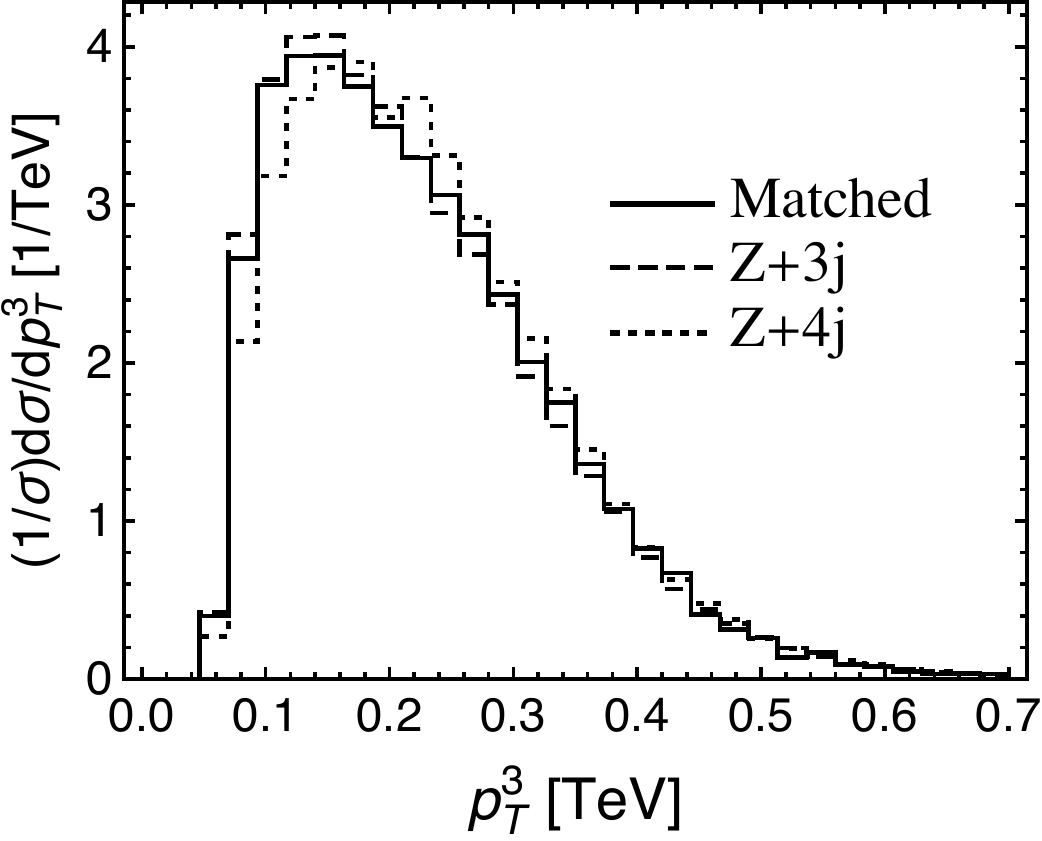}
\hfill
\includegraphics[width=0.45\textwidth]{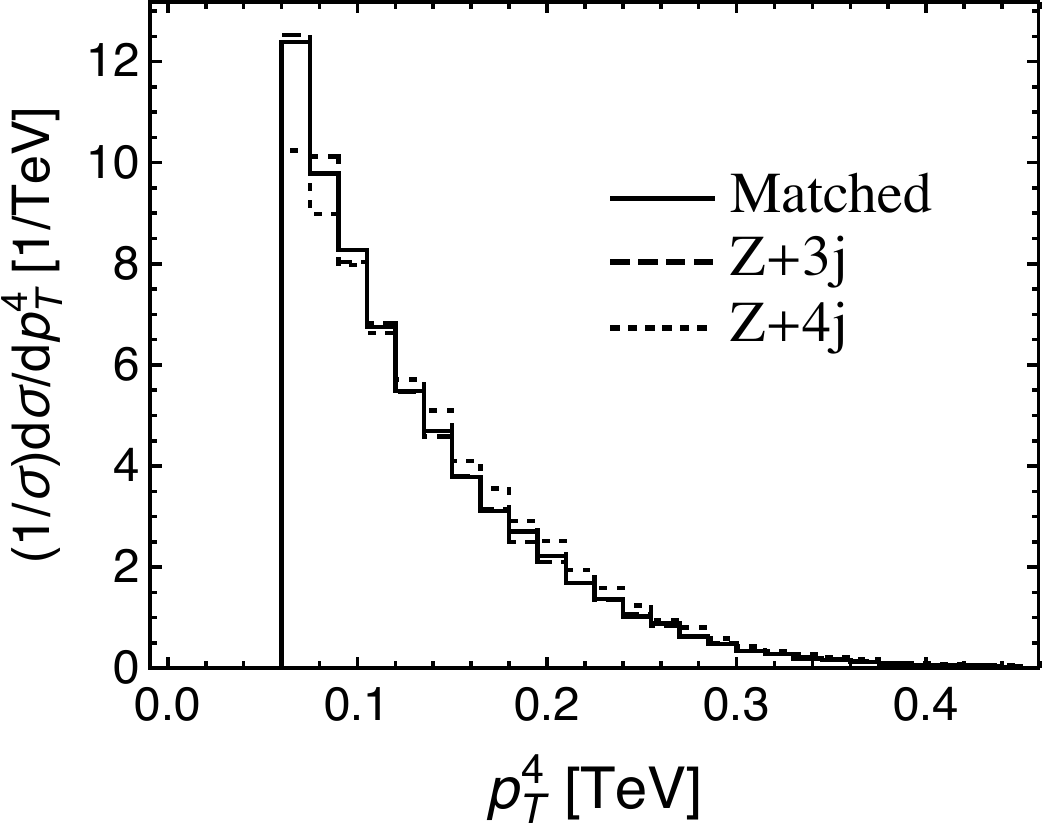}
\hfill
\caption{\label{fig:compare1}\small \sl Normalized distribution of the inclusive and exclusive kinematic variables using three different event samples: (1) $Z+$jets, ME-PS merged upto $4-$jets, (2) $Z+3-$jet ME followed by PS and (3) $Z+4-$jet ME followed by PS . The distributions are presented after {\tt Cut-1} and an $M_{\rm eff}$ cut of $1.8$ TeV, and for this reason the $H_T$ and $p_{Tj}$ distributions have a non-standard shape.}
\end{figure}

\begin{figure}[htb!]
\centering 
\includegraphics[width=0.45\textwidth]{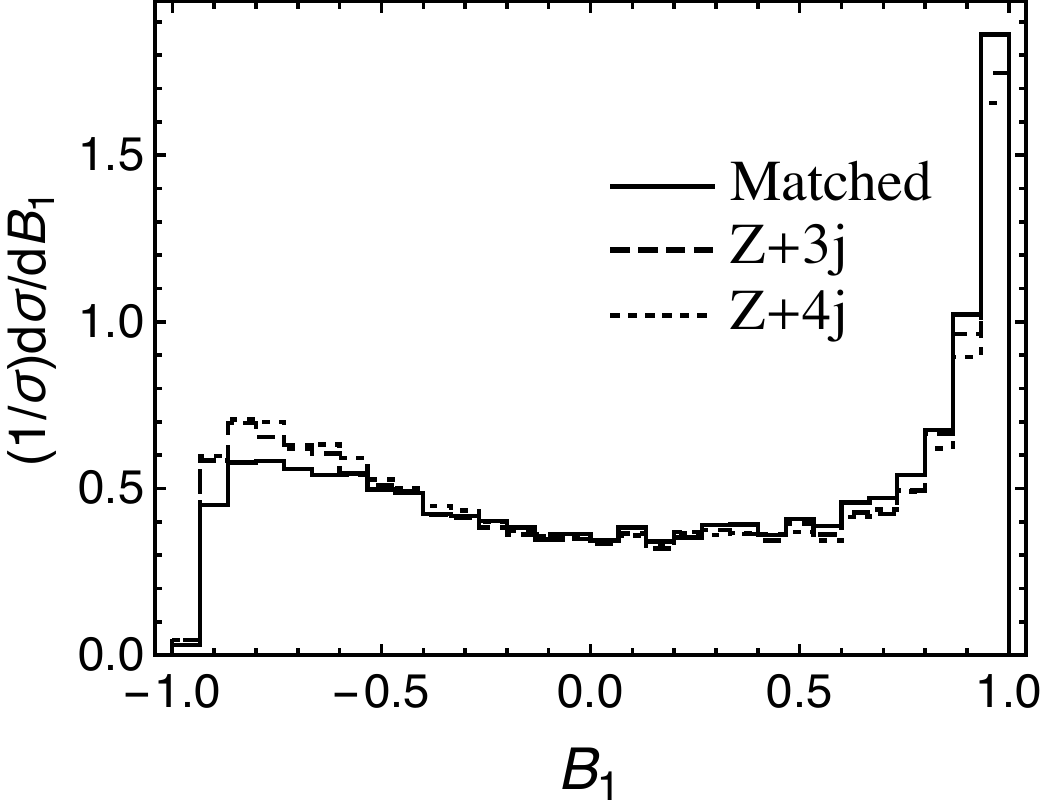}
\hfill
\includegraphics[width=0.45\textwidth]{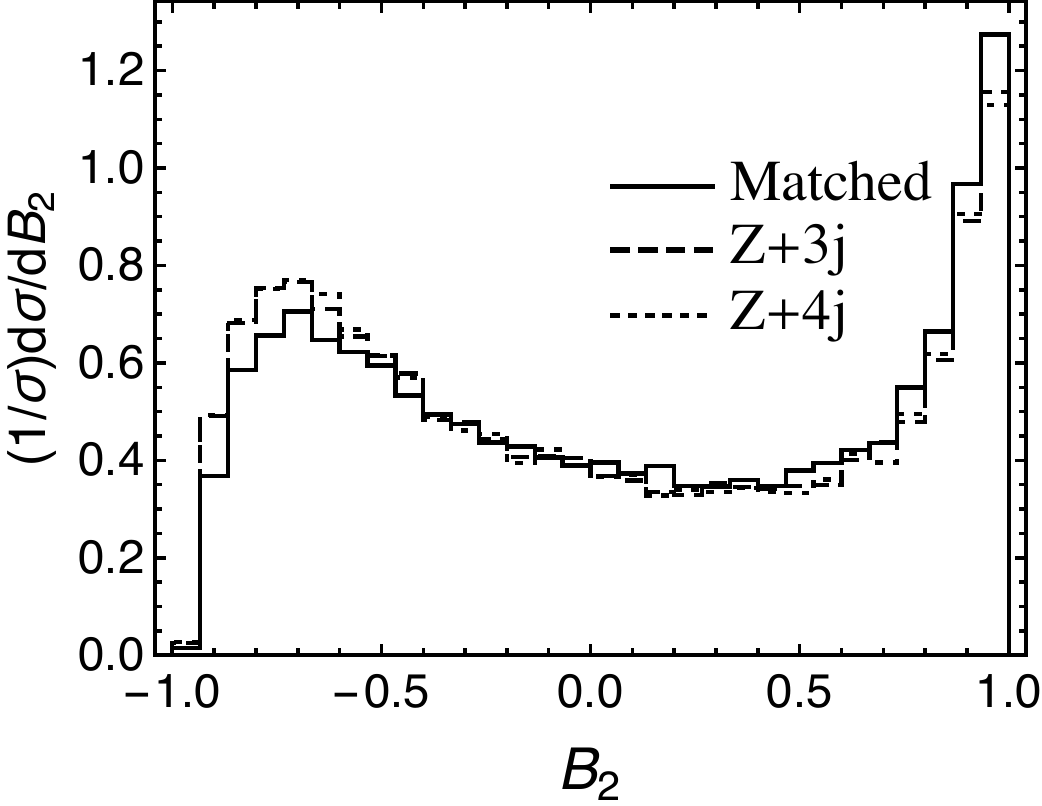}
\hfill
\includegraphics[width=0.45\textwidth]{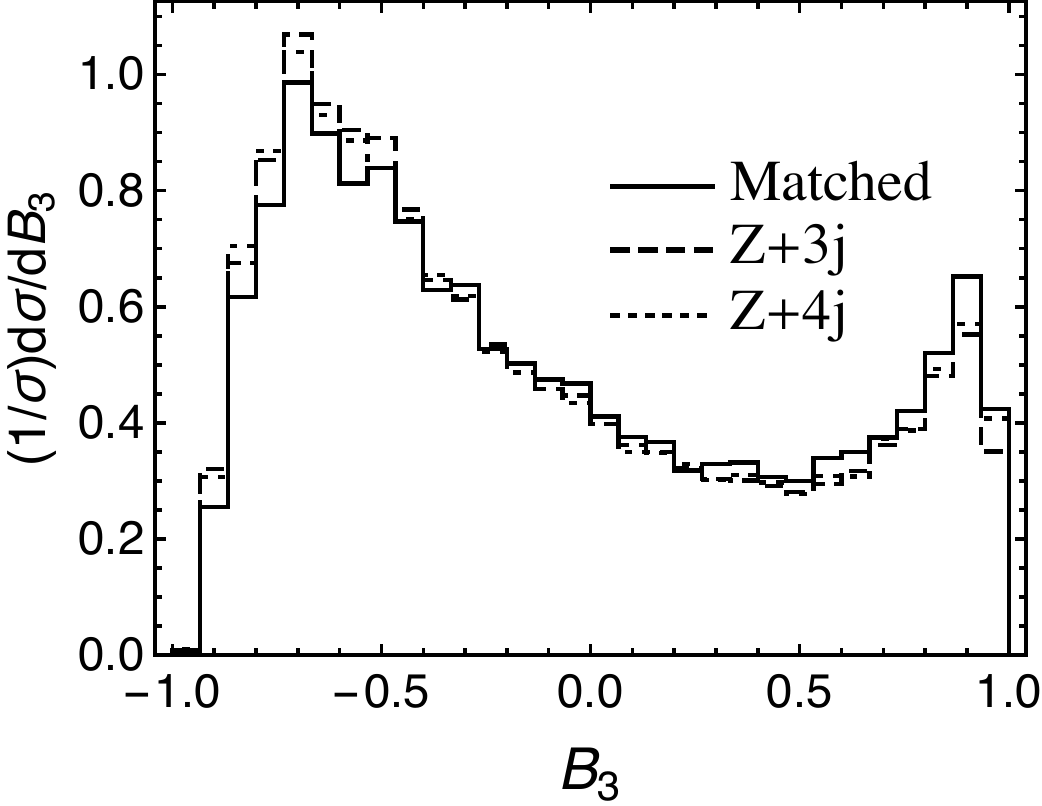}
\hfill
\includegraphics[width=0.45\textwidth]{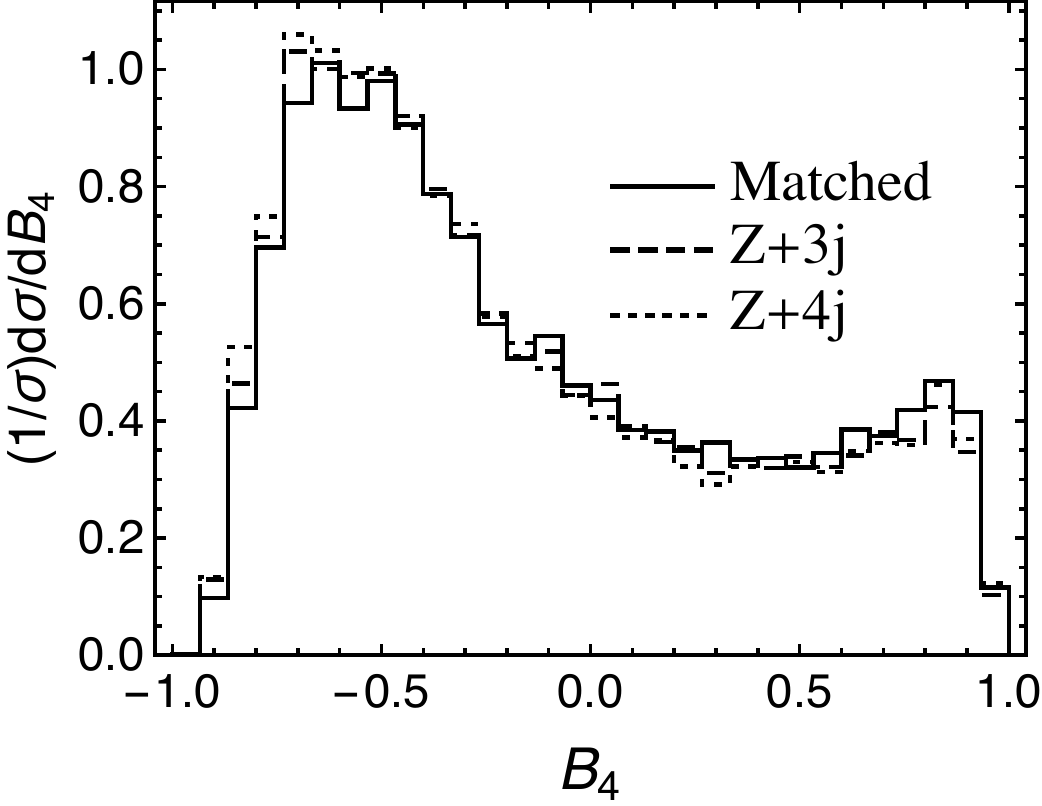}
\caption{\label{fig:compare2}\small \sl Normalized distribution of the jet substructure based BDT variables, using three different event samples: (1) $Z+$jets, ME-PS merged upto $4-$jets, (2) $Z+3-$jet ME followed by PS and (3) $Z+4-$jet ME followed by PS . The distributions are presented after {\tt Cut-1} and an $M_{\rm eff}$ cut of $1.8$ TeV.}
\end{figure}

In this appendix, we discuss the details of our simulation of the $Z+$jets background process. As discussed in Sec.~\ref{MC}, due to the necessity to generate several large statistics event samples as an input to the MVA after {\tt Cut-1} and different values of high $M_{\rm eff}$ cuts, we use the $Z+3-$jets ME (followed by PS) event sample, since it accurately reproduces normalized differential distributions for all the input variables, and is less resource intensive. The overall normalization of the $Z+$jets background is fixed by comparison with the ATLAS simulation results in the {\tt 4jm} category~\cite{ATLAS14}. This method is also cross-checked by reproducing to a good accuracy the ATLAS projected exclusion contour~\cite{ATLAS14}. In Figs.~\ref{fig:compare1} and ~\ref{fig:compare2}, we show the normalized distributions of the inclusive, exclusive and jet substructure variables utilized in this study for the three event samples of (1) $Z+$jets, ME-PS merged upto $4-$jets, (2) $Z+3-$jet ME followed by PS and (3) $Z+4-$jet ME followed by PS. As we can see from this figure, the difference in shape between these three event samples is negligibly small.

\section*{Acknowledgments}
YS is grateful to Michihisa Takeuchi for helpful discussions and computing support. The work of BB is supported by the Department of Science and Technology, Government of India, under the Grant Agreement number IFA13-PH-75 (INSPIRE Faculty Award). SM is supported in part by the U.S. Department of Energy under grant No. DE-FG02-95ER40896 and in part by the PITT PACC. M.N. and Y.S.  are partially supported by the Grant-in-Aid for Scientific Research from the Ministry of Education, Science, Sports, and Culture (MEXT), Japan (Nos.~16H06492 and 16H03991 for M.~M.~Nojiri), and also by the World Premier International Research Center Initiative (WPI Initiative), MEXT, Japan. BW is grateful for the hospitality of Kavli IPMU while part of this work was performed.

 \end{document}